\documentclass[letterpaper,twocolumn,american,showpacs,prl,aps,superscriptaddress]{revtex4}
\usepackage[latin1]{inputenc}
\usepackage{bm}
\usepackage{multirow}
\usepackage{amssymb}
\usepackage{amsbsy}
\usepackage{amsmath}
\usepackage{stmaryrd}
\usepackage{graphicx}
\usepackage{subfigure}
\usepackage{epsfig}
\usepackage{hyperref}
\makeatletter
\usepackage{pifont}
\usepackage{color}
\makeatother

\begin{document}

\title{Necessity of eigenstate thermalization for equilibration towards unique expectation values when starting from generic initial states}

\author{Christian Bartsch}
\email{cbartsch@uos.de}
\affiliation{Fachbereich Physik, Universit\"at Osnabr\"uck,
             Barbarastrasse 7, D-49069 Osnabr\"uck, Germany}

\author{Jochen Gemmer}
\email{jgemmer@uos.de}
\affiliation{Fachbereich Physik, Universit\"at Osnabr\"uck,
             Barbarastrasse 7, D-49069 Osnabr\"uck, Germany}

\date{\today}

\begin{abstract}
We investigate dynamical equilibration of expectation values in closed quantum systems for realistic non-equilibrium initial states. Thereby we find that the
corresponding long time expectation values depend on the initial expectation values if eigenstate thermalization is violated. An analytical expression for
the deviation from the expected ensemble value is derived for small displacements from equilibrium. Additional numerics for 
magnetization and energy equilibration in an asymmetric anisotropic spin-1/2-ladder demonstrate that the analytical predictions persist beyond the limits of the theory.
The results suggest eigenstate thermalization as physically necessary condition for initial state independent equilibration. 

\end{abstract}

\pacs{
05.30.-d, 
 03.65.Yz, 
05.70.Ln  
75.10.Jm 
}

\maketitle

The occurrence of statistical phenomena like equilibration and thermalization in closed quantum systems in spite of the underlying unitary dynamics receives much interest in the present research. 
One issue in this context is whether density matrices of subsystems are typically thermal \cite{lychkovskiy2010,goldstein2006,popescu2006,eisert2015,reimann2010}. 
Other investigations address the question if physical quantities like expectation values dynamically approach constant equilibrium
values that are consistent with corresponding ensemble values \cite{reimann2015,reimann2016}.
The works \cite{yurovsky2011,gogolin2011,khodja2015,khodja2016,depalma2015} indicate that equilibrium values can depend on the concrete initial state under certain conditions. 

One approach to this topic is the eigenstate thermalization hypothesis (ETH) \cite{deutsch1991,srednicki1994,rigol2008}.
The ETH implies that expectation values of an observable of interest $A$ with regard to energy eigenstates $\vert n\rangle$ vary slowly as a function of the 
corresponding eigenenergies $E_{n}$, which means that all diagonal elements of $A$ in the energy eigenbasis, i.e., $A_{nn}=\langle n\vert A\vert n\rangle$, are approximately constant within some energy regime 
which the system's state is restricted to. 
For open system situations, i.e., when the global system is divided into system plus bath, there is an alternative formulation of the ETH. If the reduced density matrices of eigenstates of the global system
$\text{Tr}_{B}\{ \vert n\rangle\langle n\vert\}$ vary slowly as a function of the global eigenenergies $E_{n}$ in terms of a suitable operator norm, then the ETH is valid. This is equivalent to the statement
that the previous definition is fulfilled for all observables acting solely on the system $A=A_{S}\otimes \mathbb{I}_{B}$.

If the ETH is fulfilled, then there is initial state independent (ISI) equilibration of expectation values with regard to $A$ \cite{rigol2008}.
If the ETH is not fulfilled, there are still initial states which yield ISI equilibration.
They even form the majority of possible initial states in a statistical sense according to the Haar measure \cite{reimann2015,hutter2013}.
However, in those cases it is usually required that the Hamiltonian, the regarded observable and the initial state are uncorrelated
meaning that the mutual orientations of 
the eigenbases of those operators can be viewed as more or less random and therefore statistically independent \cite{reimann2015,ikeda2011}. 
While this assumption is mathematically reasonable, in the sense of a high relative frequency with respect to the unitary Haar measure, its physical justification is at the heart of the present investigation.
Moreover, some investigations rely on rather strong restrictions on the initial state like identical amplitudes on all energy eigenstates in the regarded energy shell \cite{ikeda2011,riera2012}.

In this Letter we address the question whether the validity of the ETH is needed for ISI equilibration of expectation values in a realistic
setting, i.e., under realistic modeling of the  initial state. To this end we first introduce a parametrized class of initial states (\ref{inst}) and  derive an
analytical expression which quantifies the initial state dependence for small deviations
from equilibrium. Furthermore, we numerically demonstrate for various spin-based examples, that the analytically predicted small deviation behavior persists, even
for substantial deviations from equilibrium.

We investigate expectation value dynamics of an observable $A$ in a system described by some Hamiltonian $H$.
We focus on a class of initial states $\rho_{0}$, which we consider to correspond to very many realistic preparation methods, i.e.,
\begin{equation}
 \rho_{0}= \frac{1}{Z} e^{-\beta H + \delta A  }
\label{inst}
\end{equation}
with $Z=\mathrm{Tr} \{ e^{-\beta H + \delta A  } \}$ being the partition function.
The general preparation procedure is as follows. First, the system $H$ is subject to a further
potential, which is of the same form as the observable $A$. The strength of this potential is controlled
by the displacement parameter $\delta$. Then the ``shifted'' system is equilibrated into a Gibbs
state by means of a heat bath. After preparation the potential and the bath are instantly removed,
i.e., the system is quenched and consequently evolves as a closed system.
The procedure both applies when the global system $H$ is composed of a small system plus a
large bath, not to be confused with the bath used for preparation, and also when this is not the case. 

An example for the system plus large bath situation, cf. \cite{depalma2015,reimann2010}, would be a Brownian particle subject to a harmonic
potential, which is, e.g., realized by an optical trap \cite{franosch2011} . Then $H=H_{S}+H_{B}+W_{SE}$, where $W_{SE}$ describes
the perhaps complex interaction between particle and environment. The observable of interest is the 
position of the particle $A=x_{S}$, the corresponding potential term in (\ref{inst}) causes a shift
of the harmonic potential in configuration space. If $W_{SE}$ is weak, (\ref{inst}) is close to a product state,
which is a standard situation in the literature, e.g., \cite{depalma2015}.

However, the main focus of the present investigations lies on situations, where the system plus large bath decomposition does not apply. Either since the bath is not large 
or, as an extreme case, since it simply does not exist. 

An example for the latter would be a molten salt or a ionic liquid in a crucible, which is initially subject to a homogeneous electric field. The intention is to observe 
the relaxation of the spatial charge separation, i.e., the observable of interest $A$ is the dipole moment of the molten salt. Obviously, the pertinent operator 
to account for the initial homogeneous electric field is of the same dipole form. Hence the initial state may be modeled by (\ref{inst}). However, the Hamiltonian 
comprises no part on which the dipole operator $A$ does not act, hence it involves no bath. This type of situation is also illustrated by our first numerical example.

Moreover, there are settings, where initial states of a form close to (\ref{inst}) appear, even if the afore-mentioned preparation procedure is not applicable, because the
regarded observable cannot be driven out of equilibrium by 
a potential. An example would be a system of two similar sized pieces of some material, which are equilibrated at different temperatures and then brought into contact
via some weak coupling, whereupon equilibration of heat is observed. In this case the observable $A$ is the difference of the local energies of the two pieces. 
If the coupling is weak, the respective initial state (\ref{inst}) approximately assumes product form, which is the standard modeling of this situation \cite{ponomarev2011}. 
Our second numerical
example addresses this setting.
Note that the setting may not be cast into the small system plus large bath form, as addressed, e.g., in \cite{depalma2015}. 
Though less present in the literature, an equally sized subsystem situation is, e.g.,  encompassed in an 
analysis which, other than the present work, relies on Hamiltonians being drawn at random \cite{goldstein2010-1}. 

First, we treat the case of small displacements analytically.
If we assume that $\|\delta A\| _{2} \ll \|\beta H\| _{2}$, the exponential in (\ref{inst}) can be expanded using a Kubo-type relation 
from the context of linear response theory \cite{kubo1991} which yields 
\begin{equation}
 \rho_{0}=\rho_{\mathrm{eq}}(1+\frac{1}{\beta}\int_{0}^{\beta} d\lambda\ e^{\lambda  H} (A - \langle A\rangle) \delta  e^{-\lambda  H} + \mathcal{O}(\delta^2 )) 
 \label{explin}
\end{equation}
where $\langle A\rangle=\mathrm{Tr}\{ A \rho_{\mathrm{eq}}\}$ corresponds to the ``equilibrium'' expectation value with $\rho_{\mathrm{eq}} = (1/Z_{0}) e^{-\beta H}$ and $Z_{0}= \mathrm{Tr}\{ e^{-\beta H}\}$.
(In the remainder of this Letter we consider only the leading order.)
 Note that the above expansion is known to be well-controlled, i.e., it surely converges, since all expressions are analytical and all regarded operators are
assumed to have bounded spectra. Thus the truncation to linear order necessarily yields correct results for small enough deviations from equilibrium.

The matrix elements in the eigenbasis of $H$ can be evaluated as
\begin{equation}
 \rho_{0,mn} = g_{mn}(\delta_{mn} + (A_{mn}-\langle A\rangle \delta_{mn})\delta ) \ ,
 \label{instelem}
\end{equation}
with the Kronecker delta $\delta_{mn}$, 
\begin{equation}
 g_{mn}= \frac{1}{Z_{0}} e^{- (1/2) \beta (E_{m}+E_{n})} \frac{\mathrm{sinh}( \frac{\beta}{2} (E_{m}-E_{n}))}{ \frac{\beta}{2} ( E_{m}-E_{n})} \ ,
\end{equation}
and $E_{n}$ being the energy eigenvalues. Note that $g_{nn}=\rho_{\mathrm{eq},nn}$. Equation (\ref{instelem}) particularly implies that the diagonal elements of 
$\rho_{0}$ depend in detail on those of $A$.
This does not conform
with the assumption that diagonal elements of $\rho_{0}$ can be described as a smooth function of energy plus some unbiased fluctuations as in \cite{reimann2010,ikeda2011}, 
unless the diagonal elements of $A$ themselves are a smooth function of energy. The latter, however, is equivalent to the ETH being fulfilled.

In the eigenbasis representation of $H$ the initial expectation value $a(0)=\mathrm{Tr} \{ A\rho_{0}\}$ reads
\begin{equation}
a(0)= \langle A\rangle + c \,\delta \, ,  \ c=\sum_{m,n} \vert A_{mn}\vert ^{2} g_{mn} - \left(\sum_{n} A_{nn}g_{nn}\right)^2 \ ,
\end{equation}
where the coefficient $c$ takes the form of a static isothermal susceptibility which can by expressed by a Kubo scalar product $(\cdot,\cdot)$
describing the canonical correlation, $c = (A-\langle A\rangle,A )$.
If the non-resonance-condition (NRC) \cite{reimann2008} is fulfilled (as it is in most generic systems), $a(t)$ in a sense approaches, possibly for large times, 
a value
\begin{equation}
a_{\infty}= \langle A\rangle + \tilde{c}\,\delta \, , \ \tilde{c}=\sum_{n} \vert A_{nn}\vert ^{2} g_{nn} - \left(\sum_{n} A_{nn}g_{nn}\right)^2 \ ,
\end{equation}
or, equivalently, $\tilde{c} =  (\tilde{A} - \langle A\rangle, A )$, 
where $\tilde{A} $ 
represents the diagonal part of $A$ in the eigenbasis of $H$, i.e.,
$\tilde{A}_{mn} = A_{mn}\delta_{mn}$. 
If the NRC is not completely fulfilled, 
there is possibly no direct equilibration. Then  
$a_{\infty}$ corresponds to a long time average value.

The deviations from the equilibrium value $\Delta a(0):= a(0)-\langle A\rangle$ and $\Delta a_{\infty}:=a_{\infty}-\langle A\rangle$ depend linearly on the displacement parameter $\delta$. $\tilde{c}$ describes the variance of the diagonal elements $A_{nn}$ with respect to
the probability distribution given by the $g_{nn}$.
Therefore the coefficients
$c,\tilde{c}$ are always positive, since $c\geq \tilde{c}\geq 0$.
If all $A_{nn}$ are equal, i.e., the ETH is fulfilled, 
there is ISI equilibration to the expected 
equilibrium expectation value $\langle A\rangle$, since the variance $\tilde{c}$ is equal to zero and $a_{\infty}$ is independent from $\delta$.
But if the ETH is not fulfilled, i.e., $\tilde{c}$ does not vanish,
then $a_{\infty}$ in general depends on $\delta$,
i.e., on the choice of the initial state, 
thus there is no ISI equilibration, cf. \cite{yurovsky2011,gogolin2011}. Instead the expectation value typically ``sticks'' to some long time value different from $\langle A\rangle$.
We call this feature stick effect.
Or in other words, there is no ISI equilibration without the ETH being fulfilled for the generic initial states (\ref{inst}). 
This is our first analytical main result, which suggests that the ETH is not only a sufficient but also a physically necessary condition for ISI equilibration, 
and therefore the ETH appears as the driving force behind ISI equilibration. A similar result is obtained in \cite{depalma2015} for a different setting in the context of open systems.
Our central finding is also very much in line with \cite{yurovsky2011}, where ISI equilibration is found to occur if the inverse participation 
ratio of energy eigenstates with respect to eigenstates of the observable is low.
The violation of ISI equilibration can be quantified by the relative stick effect $r=\Delta a_{\infty}/ \Delta a(0)=\tilde{c}/c$ (cf. \cite{yurovsky2011}), 
which is similar to an ETH-violation parameter which was heuristically introduced in \cite{khodja2015,khodja2016}.
 Note that the result for $r$ also bares an implication in
the opposite direction: Even if the ETH is almost fulfilled, i.e., $\tilde{c}$ is very small, there still may be generically no ISI equilibration, namely if $c$ is also very small. This
likely occurs for spatial particle dynamics in many-body localized systems \cite{goldstein2015}.
Since $r$ is independent of $\delta$, one obtains a linear relation between $\Delta a(0)$ and $\Delta a_{\infty}$.
Note that in the following numerics we regard observables with $\langle A\rangle =0$ and thus $r=a_{\infty}/a(0)$. For any observable $A'$ one can readily obtain an observable $A$
with this property by $A=A' -\langle A'\rangle$. 

\begin{figure}[htb]
\includegraphics[width=8.4cm]{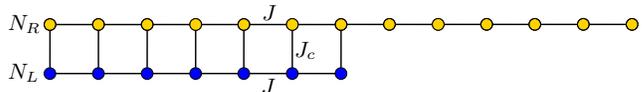}
\caption{Rotated sketch of asymmetric anisotropic (XXZ-) spin-1/2-ladder.}
\label{model}
\end{figure}

In our first numerical example we investigate the stick effect for larger displacements from equilibrium.
We consider magnetization flow in a spin system. The scenario is modeled by an asymmetric anisotropic (XXZ-)spin-1/2-ladder (sketched in Fig. \ref{model}), cf. \cite{khodja2015,khodja2016},
 \begin{equation}
   H=H_{L}+H_{R}+H_{I}
  \end{equation}
 with
 \begin{equation}
  H_{L,R}= J \sum_{i=1}^{N_{L},N_{R}} (S_{i}^{x}S_{i+1}^{x} + S_{i}^{y}S_{i+1}^{y} + \Delta S_{i}^{z}S_{i+1}^{z})
 \end{equation}
 and
 \begin{equation}
  H_{I}= J_{c} \sum_{i=1}^{N_{L}} (S_{i}^{x,L}S_{i}^{x,R} + S_{i}^{y,L}S_{i}^{y,R} + \Delta S_{i}^{z,L}S_{i}^{z,R})\ ,
 \end{equation}
where $J$ is the coupling strength along the chains, $J_{c}$ is the perpendicular coupling strength between the two chains, $\Delta$ is the anisotropy and 
$N_{L(R)}$ is the number of spins in the left (right) chain. We consider $N_L = 7, N_R =13$, i.e., a total of $20$ spins.
The left-right asymmetry suppresses trivial validity of the ETH.

First, we regard the difference of magnetization in $z$-direction between left and right chain described by the observable
\begin{equation}
   A=S_{L}^{tot}-S_{R}^{tot}- \langle S_{L}^{tot}-S_{R}^{tot}\rangle
   \label{obsmagasymm}
  \end{equation}
with $S_{L(R)}^{tot}=\sum_{i=1}^{N_{L(R)}} S_{i}^{z,L(R)}$ and $\langle\cdot\rangle = \mathrm{Tr}\{  \cdot\rho_{\mathrm{eq}}\}$.
The initial state (\ref{inst}) is then generated by applying a homogeneous magnetic field to each chain with equal strength $B$ but opposite direction, such that $\delta=\beta B$.

We focus on the  half filling subspace, i.e., total magnetization in $z$-direction equal to zero. Furthermore, we set $J=1$ and 
$\Delta = 0.1$ throughout the following investigations. We focus on two cases of weak ($J_{c}=0.2$) and strong ($J_{c}=4.5$) inter-chain coupling, which represent two particular regimes for this model
\cite{khodja2015,khodja2016}.
To keep the total energy approximately constant we fix $\beta=1$ and analyze the initial state dependence by varying $\delta$, such that the regarded initial expectation values cover the whole spectrum of $A$. 

\begin{figure}[htb]
\hspace*{-1.0cm}
\includegraphics[width=6.4cm]{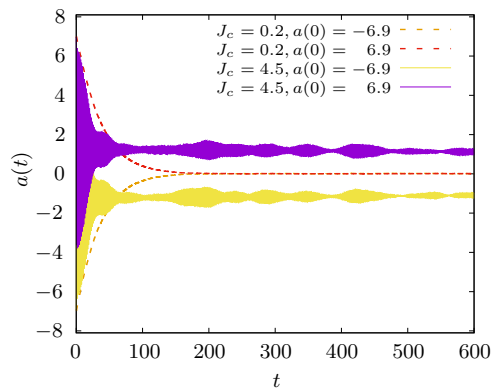}
\caption{Dynamics of magnetization difference for two different couplings $J_{c}=0.2$ and $J_{c}=4.5$ and large initial displacements. The curves for $J_{c}=4.5$ show fast oscillations and stick effect, while the curves 
for $J_{c}=0.2$ decay exponentially without stick effect.}
\label{magasymmdyn}
\end{figure}

We numerically compute the dynamics of the expectation value $a(t)=  \mathrm{Tr}\{ A \rho(t)\}$ by means of a typicality based fourth order Runge-Kutta algorithm \cite{suppmat,bartsch2009,steinigeweg2014,steinigeweg2014-1}.
This approach involves a suitable approximation of the initial state from Eq. (\ref{inst}), see also supplemental material.
Examples of the dynamics of $a(t)$ are depicted in Fig.~\ref{magasymmdyn}. 

\vspace{0.02cm}
\begin{figure}[htb]
\hspace*{-1.0cm}
\includegraphics[width=6.0cm]{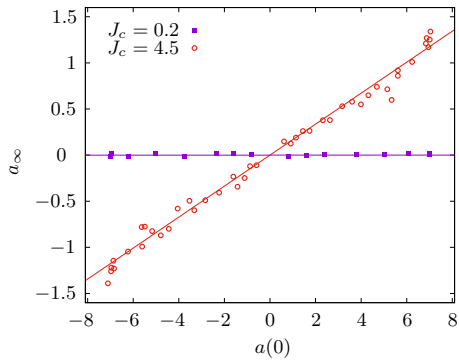}
\caption{Stick effect of magnetization difference expectation value $a$ for two different $J_{c}=0.2$ and $J_{c}=4.5$ with linear fits. 
Approximately linear dependence between $a_{\infty}$ and $a(0)$, $a_{\infty}$ is almost $0$ for $J_{c}=0.2$. Slopes of linear fits: $r=8\cdot 10^{-4}$ for $J_{c}=0.2$, $r=0.17$ for $J_{c}=4.5$.
Fluctuations are due to numerical inaccuracies.}
\label{magnumst}
\end{figure}

The weakly interacting system ($J_{c}=0.2$) shows exponential dynamics as expected, while in the case of strong ($J_{c}=4.5$) the initial dynamics are oscillating. This may be understood as dimer oscillations, since
the ladder can be viewed as chain of weakly coupled spin dimers for large $J_{c}$. Although the oscillations do not vanish completely on the regarded time scale, which may indicate that the NRC is not completely
fulfilled for this model,
those oscillations become negligibly small for large
enough initial displacements such that equilibration occurs.
The value $a_{\infty}$ is extracted from $a(t)$ by fitting a constant to the curves for large times (from $t=200$ to $t=600$). 
Figure~\ref{magnumst} shows the dependence of $a_{\infty}$ on $a(0)$ when $\delta$
is varied. 

\vspace{0.02cm}
While for the weakly interacting model ($J_{c}=0.2$) $a_{\infty}$ is almost zero through the entire spectrum, the strongly interacting case ($J_{c}=4.5$) shows a significant stick effect. 
One finds in good approximation a linear dependency of $a_{\infty}$ on $a(0)$ in the complete spectrum,
which is a simple continuation of the above analytical results for small deviations. 
The slopes of corresponding linear fit curves yield quantitative measures for the relative stick effect, one finds $r=8\cdot 10^{-4}$ for $J_{c}=0.2$ and $r=0.17$ for $J_{c}=4.5$.
Obviously, at least in these examples, the principle of no ISI equilibration 
without ETH remains valid even for large deviations.
The linear behavior is necessarily described correctly by our analytical findings. Higher orders of the expansion (\ref{explin}) cannot produce such a linear
dependence.
Note that the fluctuations of the data points in Fig.~\ref{magnumst} are numerical artifacts \cite{suppmat}.
The finding that the stick effect is much larger for the strong coupling situation is in accordance with the results in \cite{khodja2015,khodja2016} for a similar system 
and observable.

In our second numerical example we analyze energy equilibration for the same model.
In this case the regarded observable is the energy difference between left and right chain
\begin{equation}
   A=H_{L}-H_{R}- \langle H_{L}-H_{R}\rangle \ .
   \label{obsasymm}
  \end{equation}
Rather than applying an additional force term to the system the initial state is here chosen as
 \begin{equation}
 \rho_{0}= \frac{1}{Z} e^{-\beta_{L} H_{L} -\beta_{R} H_{R} -((\beta_{L}+\beta_{R})/2) H_{I} }\ ,
\label{inst2}
\end{equation}
which is of the above introduced form (\ref{inst}) with $A$ chosen
as in (\ref{obsasymm}), 
$\beta=(\beta_{L}+\beta_{R})/2$ and $\delta=-(\beta_{L}-\beta_{R})/2$. 
It is important to note that for small inter-chain couplings $H_{I}$ the last term in the exponent of (\ref{inst2}) is small and (\ref{inst2}) is approximately of the standard product form, 
which represents the setting where the left and the right chain are separately equilibrated in a canonical state (e.g., by two heat baths which are afterwards separated 
from the system) 
with inverse temperatures $\beta_{1}$ and $\beta_{2}$ and then connected via the coupling $H_{I}$.
We focus on this case in our numerical analysis and choose $J_{c}=0.2$. Larger inter-chain couplings are not considered, because (\ref{inst}) would not approximate the standard product state
any more, which would make its closeness to reality questionable.

\begin{figure}[htb]
\hspace*{-1.0cm}
\includegraphics[width=5.8cm]{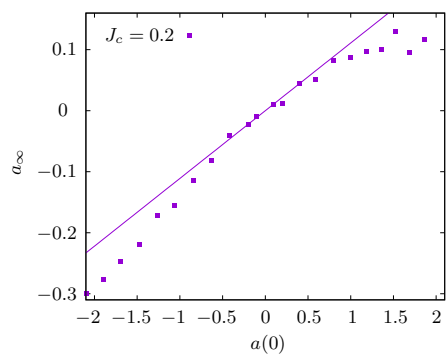}
\caption{Stick effect of energy difference expectation value for $J_{c}=0.2$  with linear fit. Approximately linear dependence between $a_{\infty}$ and $a(0)$ for $-1\leq a(0)\leq 1$. 
Slope of linear fit: $r=0.11$.}
\label{ennumst}
\end{figure}

We fix again $\beta=1$
and consider only positive temperatures, thus the reasonable regime for $\delta$ is $-\beta \leq \delta \leq \beta$.
The maximum energy difference
is realized for one of the $\beta_{L},\beta_{R}$ equal to $0$ and the inverse temperature of the colder chain equal to $2$, which corresponds to a quite low temperature,
i.e., the colder chain is mostly populated in the energy levels near its ground state. That is, we realize a substantial energy difference within this setting, although not the maximum eigenvalue of $A$.
The resulting dynamics are exponential,
Fig.~\ref{ennumst} shows the dependence of $a_{\infty}$ on $a(0)$ within the above regime.
Again, the necessarily linear relation at $a(0)=0$ extends qualitatively unaltered into a regime substantially far from equilibrium, i.e., $-1\leq a(0)\leq 1$. 
 The slope of the linear fit yields $r=0.11$.
In comparison to the first numerical example 
one finds deviations from the linear behavior for very large $a(0)$ which indicates a breakdown of the analytical predictions in these regimes.
So the validity range of our theory obviously depends on the regarded observable.

\vspace{0.02cm}
Thus, loosely speaking, even very simple non-equilibrium initial states 
like (\ref{inst2}) ``detect'' ETH violations, also beyond the small system plus large bath scenario. Note that the ETH violation occurring in this specific example may vanish
in the limit of large systems.

{\it Summary}.-
In this Letter we showed that long time equilibrium expectation values of generic non-equilibrium initial states 
deviate from the expected ensemble value
if the ETH is violated. We derived a quantitative analytical expression for this stick effect in case of small deviations from equilibrium, which implies that long time deviations scale proportional 
to the initial value. This feature was numerically verified for magnetization and energy equilibration in an asymmetric anisotropic (XXZ-)spin-1/2-ladder and 
was found to persist also for significantly large deviations from equilibrium.
Therefore, the validity of the ETH seems to be the crucial feature behind ISI equilibration for realistic initial states and may consequently be viewed as a physically necessary condition for thermalization.
This result may also be found for, e.g., fermionic or bosonic models, and may also be interesting in the context of many-body localization \cite{pal2010}, where dynamical equilibration is of
interest.

We thank D. Schmidtke for fruitful discussions.

\end{document}